# Quasar Populations in a Cosmological Constant Dominated Flat Universe

Sangeeta Malhotra and Edwin L. Turner

Princeton University Observatory, Peyton Hall, Princeton, NJ 08544

I: san@astro.princeton.edu and I: elt@astro.princeton.edu

## ABSTRACT

Most physical properties derived for quasars, as single entities or as a population, depend upon the cosmology assumed. In this paper, we calculate the quasar luminosity function and some related quantities for a flat universe dominated by a cosmological constant $\Lambda$ ($\Lambda = 0.9, \Omega = 0.1$) and compare them with those deduced for a flat universe with zero cosmological constant ($\Lambda = 0, \Omega = 1$). We use the AAT quasar survey data (Boyle et al. 1990) as input in both cases. The data are fit well by a pure luminosity evolution model for both the cosmologies, but with different evolutionary parameters. From the luminosity function, we predict (extrapolate) a greater number of quasars at faint apparent magnitudes (twice the number at B=24, $z < 2.2$) for the $\Lambda$ dominated universe. This population of faint quasars at high redshift would result in a higher incidence of gravitational lensing. The total luminosity of the quasar population and the total mass tied up in black hole remnants of quasars is not sensitive to the cosmology. However, for a $\Lambda$ cosmology this mass is tied up in fewer but more massive black holes.

*Subject headings:* cosmology, cosmological constant, quasars, massive black holes, gravitational lenses

## 1.  Introduction

Cosmological models in which a vacuum energy term $\Lambda$ dominates the total mass density term $\Omega$ to produce zero spatial curvature are consistent with the fundamental prediction (k=0) of inflationary cosmology (Guth 1981) and seem better able to accommodate many available cosmological observations (Peebles 1993, Efstathiou, Bond & White 1992, Turner 1991, Efstathiou, Sutherland & Maddox 1990, Kofman Gnedin & Bahcall 1993, Stompor & Gorski 1994). Although these models are significantly constrained by gravitational lens statistics (Maoz & Rix 1993, Kochanek 1993, Fukugita & Turner 1991) and implausible from a fundamental theoretical perspective (Weinberg 1989), these strengths justify continued attention and exploration. See Carroll, Press and Turner (1992, hereafter CPT) for a recent review of these issues.

Systematic surveys of the quasar population have now progressed to the point that direct determination of the luminosity function and related quantities has replaced parameter fitting to assumed analytic forms (see Hartwick and Schade 1990 for a recent review). The astronomical and astrophysical implications of the quasar luminosity function can therefore be considered with more confidence. For example, the long standing alternative of density or luminosity evolution has been largely resolved in favor of the latter.

In this paper we exploit modern data to investigate the implications a non-zero cosmological constant for the quasar population, subject to the k=0 constraint required by conventional theoretical blinders. The two cosmological models considered are a matter dominated flat universe (model O): $\Lambda = 0$, $\Omega = 1$, and a $\Lambda$ dominated flat universe (model L) $\Lambda = 0.9$, $\Omega = 0.1$. The particular values of $\Lambda$ and $\Omega$ are illustrative only, and are being used to emphasize the contrast between the two models. The value of Hubble's constant is assumed to be $H_0 = 50\,\mathrm{km\,s^{-1}\,Mpc^{-1}}$ to facilitate direct comparison with the luminosity function derived for model O by Boyle et al. 1987(BFSP), 1988(BSP).

## 2. Cosmology

The effect of a nonzero cosmological constant on the general quantities in the cosmologies: distance, lookback time, comoving volume etc. are well known and well studied. The derivations are given in Weinberg 1972, Peebles 1984, Charlton and Turner 1987, and CPT. Here we recapitulate some of the relations and apply them to derive qualitatively the properties of the quasars.

In a Friedman-Robertson-Walker (FRW) cosmology with zero curvature the scale factor at any epoch is given by the Friedman equation,

$$H^2 = \left(\frac{\dot{a}}{a}\right)^2 = \frac{8\pi G}{3}\rho + \frac{\lambda}{3} \tag{1}$$

Using the definitions

$$\Omega = \frac{8\pi G \rho}{3H_0{}^2}; \quad \Lambda = \frac{\lambda}{3H_0{}^2}; \quad \Omega + \Lambda = 1, \tag{2}$$

$$H^2 = \left(\frac{\dot{a}}{a}\right)^2 = H_0^2\left(\Omega a^{-3} + \Lambda\right). \tag{3}$$

The look-back time for an object at redshift z is the time taken for the photon emitted by the object to travel and is given by the expression

$$t = \int_{a(t)}^{a_0} \frac{da}{\dot{a}} = \int_{(1+z)^{-1}}^{1} \frac{da}{a(\Omega a^{-3} + \Lambda)} = \frac{2}{3} \frac{H_0^{-1}}{\sqrt{1-\Omega}} \sinh(\Omega^{-1} - 1) \qquad (4)$$

The proper distance $d_0$ to an object at redshift z is defined as the distance covered by a photon emitted by the object.

$$d_0 = H_0^{-1} \int_{(1+z)^{-1}}^{1} \frac{da}{a^2(\Omega a^{-3} + \Lambda)} \qquad (5)$$

The luminosity distance $d_L$ is defined as

$$d_L = \left(\frac{\mathcal{L}}{4\pi \mathcal{F}}\right)^{\frac{1}{2}}, \qquad (6)$$

where $\mathcal{L}$ is the intrinsic luminosity of the source, and $\mathcal{F}$ is the measured flux. The luminosity distance to an object at redshift z is given by $d_L = (1+z)d_0$. The luminosity distance corresponding to redshift z=2 is a factor $\sim 2$ larger in model L than in model O. This translates into a decrease in the absolute magnitude of $\sim 1.5$ for a quasar of the same apparent magnitude.

The comoving volume at redshift z is given by

$$dV_0(z) = \frac{1}{3}\frac{d}{dz}[d_0{}^3]dz d\Omega. \qquad (7)$$

The comoving volume is quite sensitive to the value of $\Lambda$. It peaks at $z \sim 1$ for model O ($\Lambda = 0$) and at $z \sim 2$ for model L. Also the peak value is larger for model L by a factor $\sim 5$ compared to model O.

## 3. Evolution Models

From their survey of optically selected quasars (BFSP, BSP) derive a pure luminosity evolution model of quasars. We use their data to calculate the luminosity function for the two cosmological models. Following BFSP and BSP we divide the data into four redshift bins in the interval $z = 0.3 - 2.2$ and calculate the luminosity function for each bin.

Assuming a spectrum of the form $f_\nu \propto \nu^{-\alpha}$ with the spectral index $\alpha = 0.5$, the absolute magnitude M of a quasar is given by: $M = m - 5\log(d_L/10\,\text{pc}) + (1+\alpha)\log(1+z)$.

The main effect of the cosmology is to make the intrinsic luminosity of the quasars higher for the same apparent magnitude in the $\Lambda$ dominated model, this effect being larger for higher redshift.

The luminosity function $\Phi(z, M_b)$ for the redshift bin $\Delta z$ and magnitude bin $\Delta M$ is given by

$$\Phi(M, z) = \sum_{\substack{j \\ z_j \in (z-\frac{\Delta z}{2}, z+\frac{\Delta z}{2}) \\ M_j \in (M-\frac{\Delta M}{2}, M+\frac{\Delta M}{2})}}^{n} \left(V_a^j\right)^{-1} \quad (8)$$

with an error on $\Phi$

$$\Delta \Phi = \left(\sum_j^n V_a^{j^{-2}}\right)^{1/2} \quad (9)$$

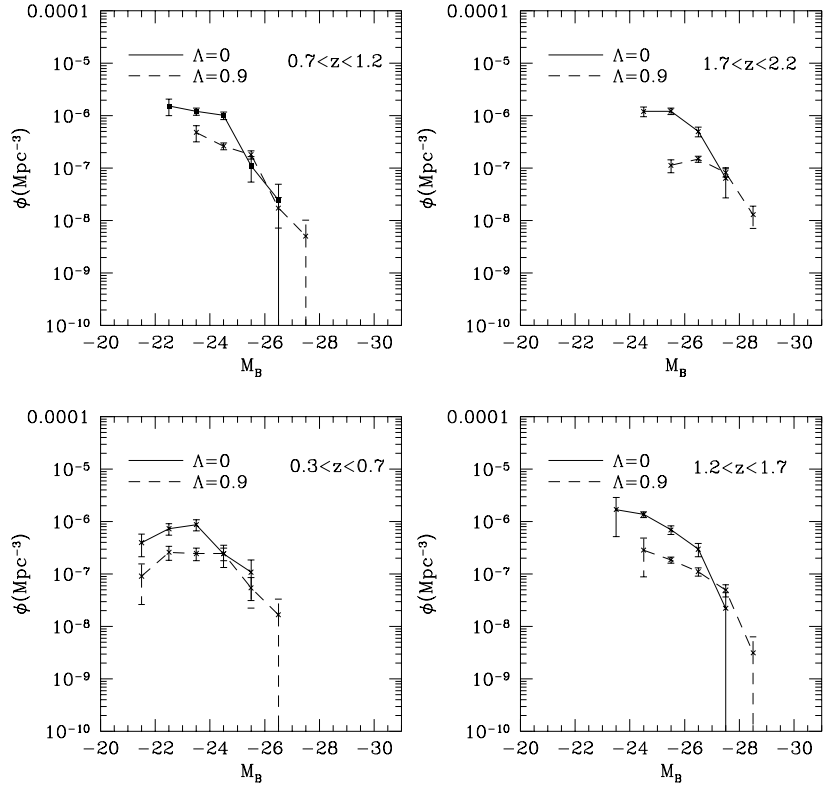

**Figure 1:** Luminosity function for the two models for different redshift bins. Model O (*solid lines*) and Model L (*dashed lines*).The binned luminosity function is given by equations (9) and (10). The luminosity functions for both the cosmological models are qualitatively similar.

Where $V_a$ is the accessible comoving volume (Avni and Bahcall 1980). We see that the luminosity function for the two models as shown in Figure (1) can be described by a two power law fit, with a shallower slope at low luminosities, for the model L as well for model O (also derived by BSP). The luminosity function curve is shifted down and to the right for the $\Lambda$ model. This is because the inferred luminosity and the comoving volume are higher in model L. The difference between the two models is greatest in the highest redshift bin $(1.7 < z < 2.2)$.

Since the luminosity functions look similar for both the cosmological models, and the luminosity function in the $\Lambda = 0$ model is well fit by a two power law with the characteristic luminosity $L_0$ evolving as a power law with redshift (equations 10 & 11) (BFSP); we fit the same model in the case of the $\Lambda$ cosmology.

$$\Phi(M,z) = \frac{\Phi^*}{10^{0.4(M-M_z)(\alpha+1)} + 10^{0.4(M-M_z)(\beta+1)}} \qquad (10)$$

where the characteristic magnitude (the knee of the two power law distribution) shows an evolution with z.

$$M_z = M_0 - 2.5 k_L log(1+z) \qquad (11)$$

The best fit luminosity function is derived by maximizing the likelihood function (given by Marshall et. al. 1983); or minimizing $S = -2lnL$. For a complete sample

$$S = -2\sum_{i=1}^{N} ln[\Phi(M_i, z_i)] + 2\int_{z_1}^{z_2}\int_{M_{lim}(z)}^{-\infty} \Phi(M,z)\Omega(M,z)\frac{dV_0}{dz}dzdM \qquad (12)$$

where $\Omega(M,z)$ is the area covered by the survey and $M_{lim}(z)$ is the faintest magnitude observable at redshift z. The minimization is done using a downhill simplex routine, AMOEBA (Press et al. 1993).

Confidence intervals around the best fit values were estimated by calculating $\Delta S$, varying two parameters at a time. For two parameters 68% levels correspond to $\Delta S = 2.3$ (Lampton, Margon & Bowyer 1976). This also enabled us to see the correlation of errors in parameter values. Errors on the pair of parameters $\beta$ and $k$ are found to be correlated. The slope of the bright end of the quasar luminosity function ($\alpha$) is ill-constrained by this sample but can be derived from the Bright Quasar Survey (Schmidt & Green 1983). Uncertainty in this parameter does not affect the results in the rest of this paper, which are sensitive to the faint end of the quasar luminosity function.

The best-fit models were tested for goodness-of-fit using the 2-dimensional K-S statistic (Press 1993, Peacock 1983) and the $\chi^2$ test for the binned luminosity function:

$\chi^2 = (\Phi - \Phi_{model})^2$; where $\Phi_{model}$ was calculated by averaging the best fit luminosity function over the redshift and magnitude bins. There is a tendency for the luminosity function of highest magnitude bin to be slightly lower than expected by the model. This is due to the coarse size of the bins. The faintest magnitude bins are not completely within the flux limit. We correct for this affect by considering only the part of the bin that is detectable in the survey.

Table 1 shows the results of the KS and the binned $\chi^2$ tests as well as the best fit parameters. The KS probability for the model L is lower than that of model O, whereas the $\chi^2$ test shows a slightly higher acceptability for model L. Both fits are deemed acceptable.

## 4. Number-Magnitude relationship

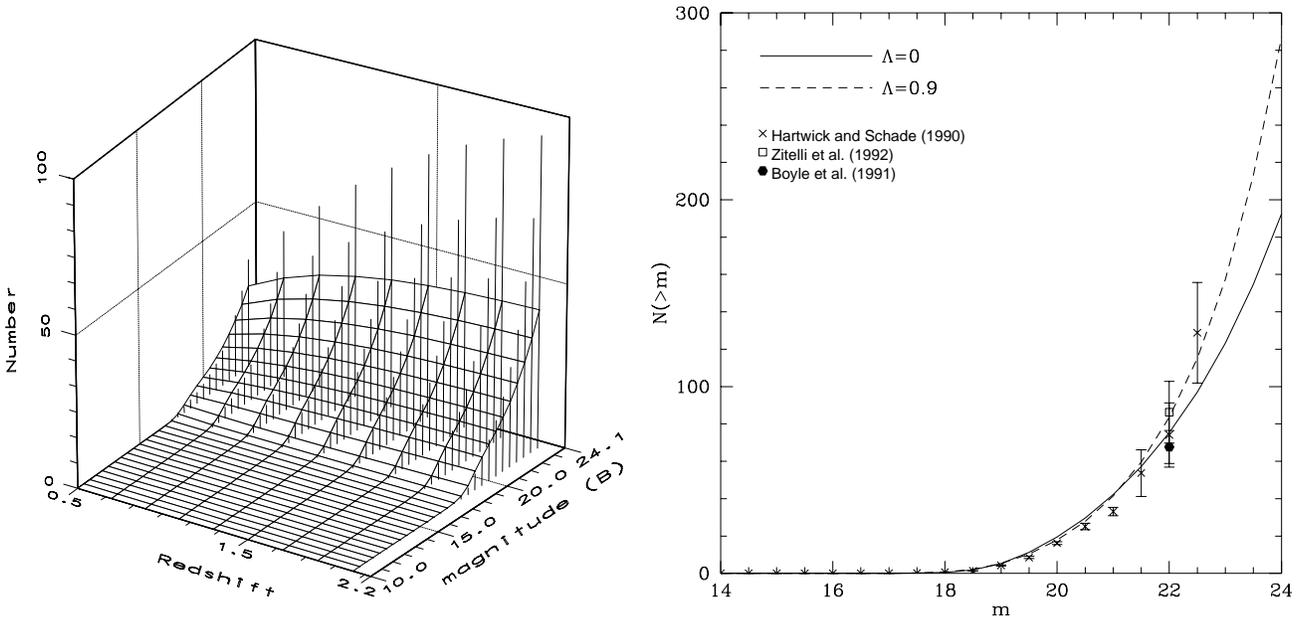

**Figure 2a:** The calculated number of quasars as function of redshift z and apparent magnitude B, Model O (*mesh*) and Model L (*spikes*). **Figure 2b:** Cumulative surface density Quasars ($deg^{-2}$) brighter than magnitude m, Model O (*solid lines*) Model L (*dashed lines*), compared with the observed surface densities (Hartwick and Schade 1990 (*crosses*) Zitelli et al. 1992 (*hollow square*) Boyle, Jones & Shanks (1991) (*hexagon*))

The best-fit luminosity models were then integrated over volume and luminosities to obtain the number of quasars per apparent magnitude for $z < 2.2$. Figure 2(a) shows how the quasars are distributed in luminosity and redshift for the two cosmological models, and Figure 2(b) shows the cumulative number-magnitude N(m) relation for the two cosmologies as well as observed N(m) from Hartwick and Schade (1990). The number of faint quasars is higher for the $\Lambda$ cosmology. This is simply due to the higher comoving volume at high ($z \simeq 2$) redshifts for this model. The observed surface densities (Hartwick and Schade 1990) as seen in Figure(2), do not as yet unambiguously distinguish between the two models.

The number of faint quasars depends most sensitively on the slope of the faint luminosity end of the spectrum $\beta$ and $k$, the evolution parameter for the characteristic luminosity. Since these parameters are also seen to be correlated, the uncertainties in the predicted number of quasars are calculated from the joint confidence intervals of these parameters. Table 2 lists the cumulative surface density of quasars, with the 1-$\sigma$ confidence interval. Better determination of the parameters $\beta$ and $k$ are needed to reduce the uncertainty in N(m).

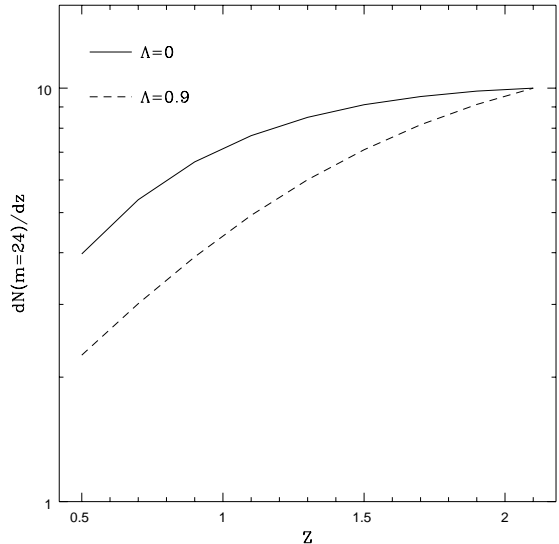

**Figure 3:** The redshift distribution of faint quasars ($24 \geq m \geq 23.5$) for the two cosmologies. The actual numbers have been normalised to facilitate comparison.

The faint quasars are also predicted to have different distributions in redshift depending on the cosmological model. Figure 3 shows the distribution of 24th magnitude quasars in redshift. The counts have been normalized to facilitate comparison of profiles. This test, combined with the number-magnitude prediction allows a more robust test, in that one can distinguish between cosmological effects from, for example, a different population of intrinsically faint quasars at low redshifts swelling the numbers at faint magnitudes.

## 5. The Number of Lensed Quasars

Besides depending upon the number density of galaxies and their properties, the number of quasars gravitationally lensed also depends upon the number of unlensed quasars. Because of the amplification bias, i.e. the brightening of lensed quasars, the number of quasars a few magnitudes fainter than the detection limit is important in predicting the number of lensed quasars in a flux limited sample. Since the number of faint quasars is greater for the $\Lambda$ dominated cosmology, and the excess faint quasars are at high redshifts, the number of lensed quasars is expected to be higher. This is in addition to the larger optical depth of the $\Lambda$ dominated universe (Turner 1990, Fukugita et al. 1990).

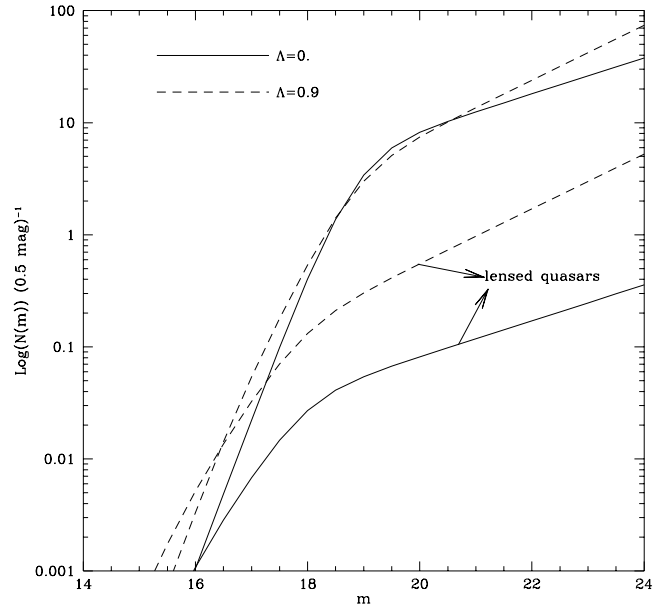

**Figure 4:** The (differential) number-magnitude relation for quasars in the two cosmologies. The number of quasars per half-magnitude bins is plotted for the model O (*solid lines*) and Model L (*dashed lines*). The number-magnitude relation for the number of lenses expected is also plotted. This calculation takes into account amplification bias, and assumes that the luminosity function is not affected by lensing (i.e. most of the observed quasars are unlensed). This assumption may break down at the bright end (m=15) of the luminosity function

Fukugita & Turner (1991, FT) calculated the number of lensed quasars expected for $\Lambda$ dominated cosmology taking into account the amplification bias. Here we repeat their calculations taking into account the luminosity function for quasars in this cosmology.

Assuming that a small fraction of observed quasars are lensed and that the probability of of lensing is independent of the absolute luminosity of the quasar, the number-magnitude counts of the lensed quasars are given by the relation (FT)

$$N_{LQ}(m) = \tau_{GL} \int_0^\infty N_Q(m+\Delta)P(\Delta)d\Delta \qquad (13)$$

where $P(\Delta)d\Delta$ is the probability that lensing will cause the magnitude of the quasar (images) to decrease by $\Delta$; i.e. the image(s) brighten by a factor A: $\Delta = 2.5\log A$. The amplification probabilities are the SIS point source amplification probabilities (Turner, Ostriker & Gott 1984, hereafter TOG). $P(\Delta)d\Delta = 7.3710^{-0.8\Delta}d\Delta$.

The number-magnitude relation for lensed quasars is then given by

$$N_{LQ}(m) = \int_0^\infty d\Delta P(\Delta) \int_{z1}^{z2} dz \tau_{GL}(z) \frac{dV}{dz} \int_{M_1(z)}^{M_2(z)} \Phi(M,z)dM \qquad (14)$$

The optical depth to lensing $\tau_{GL}(z)$ is given by (Turner 1990)

$$\tau_{GL}(z) = \frac{F}{30}\left[\int \frac{dw}{(\Omega_0 w^3 - \Omega_0 + 1)^{1/2}}\right]^3 \qquad (15)$$

The parameter F depends on the number density of galaxies and their effectiveness in lensing. $F = 16\pi^3 <n_0\sigma_\parallel^4>/cH_0^3$. From the properties of local galaxies, FT estimate this to be: F=0.047. Using this value in our calculations, the number of lensed quasars expected in the BFSP sample for both the cosmological models is plotted in Figure 4. For the BFSP sample with a magnitudes limit 20.9, we expect $\simeq 0.5$ lensed quasars for $\Lambda = 0, \Omega = 1$ cosmology, and $\simeq 3.5$ lensed quasars for $\Lambda = 0.9, \Omega = 0.1$. These estimates are slightly smaller than given by Turner (1990), but larger than the numbers estimated by FT. This is because we use the reduced value of lensing cross-section of galaxies ($F$) estimated by FT, but take into account the large number of faint quasars in the $\Lambda$ cosmology, which leads to an increase in the expected number of lensed quasars.

## 6. Black Hole Remnants

Since the cosmological constant has the affect of increasing the luminosity distances and hence the absolute luminosity of observed sources, more massive black hole remnants are implied. Assuming the quasars to be powered by accretion on to black holes one can calculate the mass density of the remanents from the the integrated luminosity density of all the quasars and assuming some efficiency of mass-energy conversion $\epsilon$ (Soltan 1982, Chokshi & Turner 1992)

$$\rho_{BH} = \int \int \epsilon L \Phi(L,z) dL \frac{dt}{dz} dz \qquad (16)$$

The total mass density in the black hole remanents is independent of the cosmological model if we consider only the observed quasars and not the predicted rise in the number of faint (and as yet unobserved) quasars. A simple argument to why this should be is as follows. For the same number and apparent magnitude of objects whose luminosity $\propto d_0^2$, density $\propto d_0^{-3}$ and time $\propto d_0^1$, the luminosity density integrated over time is independent of distance and hence of the cosmological model used. Including extrapolations to fainter populations gives a higher value of $\rho_{BH}$ in the $\Lambda$ cosmology by only $\sim 20\%$. This is because faint magnitude (m$>$ 21) quasars contribute little to the luminosity density budget.

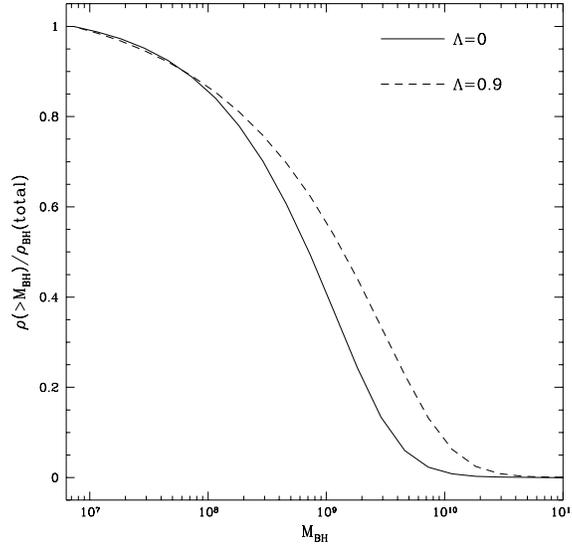

**Figure 5:**The cumulative mass per comoving volume contributed by black hole remnants of mass $M_{BH}$ in the two cosmological models. Pure luminosity evolution models is assumed for individual quasars. The total mass in black hole remnants differs by less than 20% in the two models, but in model L the mass is tied up in bigger black holes.

The arguments given above also show that the luminosity density of the quasars is not very different in the two models so the effect on the intergalactic medium will be similar.

One of the slightly more 'natural' explanations for the observed luminosity evolution is to have a population of quasars luminous at $z \simeq 2$, with the individual quasars fading to the present epoch. The luminosity evolution of the quasars could also be explained as the evolution of a population of quasars whose mean luminosity is decreasing from $z \simeq 2$ to the present. The first scenario would result in few, but massive blackhole remanents, whereas the second scenario produces a larger number of modestly sized BHs.

The pure luminosity evolution model of individual quasars would lead to more massive, but fewer (in number) black holes for a $\Lambda$ dominated cosmology than otherwise. This is because we infer higher intrinsic luminosity for the quasars in that cosmology and also because the lookback time is longer. The contributions to $\rho_{BH}$ from different quasar luminosities and hence different black-hole masses (assuming that quasars do not exceed their Eddington luminosity (Rees 1984, Jaroszynski 1980)) is different. We find that for model O there is negligible contribution from black holes with mass $M_{BH} \geq 10^{10} h^{-2} M_\odot$ whereas for model L $M_{BH} \geq 10^{10} h^{-2} M_\odot$ contribute 10% to the total $\rho_{BH}$ (Fig(5)). The number density of such objects is $\sim 3$ times higher for $\Lambda$ dominated model L.

## 7. Discussion

The quest to determine the parameters that characterize the FRW universe has been a long and difficult one. One of the main difficulties is the separation of intrinsic properties and evolution of the objects studied from the cosmology assumed. To make statements about one, one has to make some assumptions about the other. This is particularly true for quasars which are found at high redshifts and are seen to evolve rapidly with redshift.

Recent surveys (BFSP, BSP) characterize the luminosity function of the quasars at faint magnitudes (m < 20.9, z < 2.2), and show that a two power law model with pure luminosity evolution fits the observations well. This is assuming FRW cosmology with q=0 or 0.5, and $\Lambda = 0$. We show in this paper that for a flat universe with a cosmological constant $\Lambda = 0.9$, the quasars also show pure luminosity evolution. The small density evolution postulated for faint quasars in this sample by BSP is not necessary in the $\Lambda$ dominated cosmology. The Pure Luminosity Evolution (PLE) of quasars allows us to explore the effects of the different comoving volumes in the different cosmologies. Comparing a $\Lambda$ dominated flat universe to a flat universe with $\Omega_0 = 1$, the most dramatic difference lies in the comoving volume $dV/dz$. This leads to a higher number of faint objects. Indeed this feature was used to explain the excess in the number counts of faint blue galaxies (Fukugita et al. 1990), but no such excess was seen in the near-IR K band counts (Gardener, Cowie & Wainscot 1993). Loh & Spillar (1986) used PLE assumption for galaxy counts to put constraints on $\Omega$ (but see Bahcall & Tremaine 1988). Recently Schade & Hartwick (1994) used the X-ray selected quasars and PLE to apply the Loh-Spillar test to rule out $\Omega = 1$.

Exploiting the pure luminosity evolution of quasars, borne out by the data for B<20.9, z < 2.2 (BSP), we extrapolate the number-magnitude relation to 24th magnitude and predict twice as many B=24 quasars in the $\Lambda$ dominated universe, as in the $\Omega_0 = 1$ model. The data at faint magnitudes, B=22.5 (Hartwick & Schade 1990, Boyle, Jones & Shanks 1991, Zitellii et al. 1992) marginally favor the $\Lambda = 0.9$ model, but are consistent with the other model as well. Better determination of the slope at the faint end (B $\simeq$ 21) of luminosity function along with a deep survey (to B=24) could help discriminate between the two cosmologies.

The number-magnitude relation for quasars is a valid test for cosmology only as long as density evolution is believed to be unimportant. We see that the number density evolution is at best negligible for all the FRW cosmologies; at least for the observed (B $<$ 20.9) quasars. Some recent studies have called into question the PLE model of quasar evolution (Hewett, Foltz & Chaffee 1993, Goldschmidt et al. 1992). Change in the number density of quasars with redshift can be attributed to both density evolution and a different $dV/dz$. Even though the form of $dV/dz$ as a function of z is fairly restricted in a FRW universe it would probably take a more subtle analysis to distinguish between different cosmological models if density evolution is also present. Both the predictions for and the observations refer to quasars at redshifts $z < 2.2$.

The larger number of faint quasars at high redshifts also lead to a prediction of a larger number of lensed quasars for model L, even in the present magnitude limited samples. This is in addition to the higher optical depth expected for $\Lambda$ dominated cosmologies. This is because of amplification bias, and the number of lenses is higher by a factor $\simeq 2$ for the BFSP sample (limiting magnitude B= 20.9) than if we had not taken into account the change in luminosity function due to the different cosmologies.

There are various uncertainties in estimating the numbers of lensed quasars ($N_{LQ}$) expected in different cosmological models due to uncertainties in the properties, numbers and evolution of lensing galaxies, but it is difficult to reconcile a $\Lambda$ dominated cosmology with the observed number of lenses (Mao 1991, Mao & Kochanek 1992, Rix et al. 1994, Fukugita & Peebles 1994). Independent of the uncertainties in the properties of the lensing galaxies, the higher populations of faint quasars would contribute to increase the $N_{LQ}$ for high $\Lambda$ cosmologies.

The other properties of the quasar (populations) affected are the luminosities, volume density, and the evolution time scale. For the $\Lambda$ model of cosmology the quasars are intrinsically brighter (since they are further away), they are also rarer (since the comoving volume is larger), and they have more time to evolve. The evolution of quasar population in time is less rapid in a $\Lambda$ dominated universe. These changes affect the expected remnants of quasars, surviving as dead black holes. The total mass in the black hole remnants is fairly insensitive to the cosmology, increasing by only 20% for the $\Lambda$ model. The mass spectrum of black holes however is different. The difference predicted by the two cosmologies however is swamped by our lack of knowledge about whether the PLE is the evolution of the population of quasars or it is the individual quasars that dim and produce the luminosity evolution we see.

## 8. Acknowledgements

We would like to thank Rachel Webster for providing computer readable version of the data and to David Spergel and Changbom Park for discussions. This work was supported by NASA grant NAGW-2173.

---

TABLE 1

BEST FIT PARAMETERS

| Model | $M_0$ | k | $\alpha$ | $\beta$ | $P(>\chi^2)$ | $P_{ks}$ |
|---|---|---|---|---|---|---|
| $\Lambda = 0 \quad \Omega = 1$ | -22.5 | 3.47 | -4.33 | -1.40 | 0.46 | 0.50 |
| $\Lambda = 0.9 \; \Omega = 0.1$ | -24.3 | 3.02 | -4.39 | -1.61 | 0.56 | 0.06 |

TABLE 2

CUMULATIVE SURFACE DENSITY OF QUASARS

| | $N(>m)$ | |
|---|---|---|
| $m$ | $Model A$ | $Model B$ |
| 19.0000 | $5^{+1}_{-0}$ | $5^{+0}_{-0}$ |
| 19.5000 | $11^{+1}_{-0}$ | $10^{+0}_{-0}$ |
| 20.0000 | $19^{+1}_{-1}$ | $18^{+0}_{-0}$ |
| 20.5000 | $30^{+2}_{-1}$ | $28^{+1}_{-1}$ |
| 21.0000 | $42^{+4}_{-3}$ | $42^{+4}_{-3}$ |
| 21.5000 | $57^{+8}_{-7}$ | $60^{+7}_{-7}$ |
| 22.0000 | $75^{+13}_{-12}$ | $84^{+13}_{-14}$ |
| 22.5000 | $97^{+20}_{-19}$ | $115^{+21}_{-24}$ |
| 23.0000 | $123^{+28}_{-32}$ | $157^{+34}_{-40}$ |
| 23.5000 | $155^{+40}_{-48}$ | $213^{+52}_{-64}$ |
| 24.0000 | $193^{+55}_{-71}$ | $287^{+78}_{-101}$ |